\documentstyle[times,pramana,epsf,floats]{ias}

\newcommand{\Br}{{\rm Br}}
\newcommand{\mtbfit}{M_{tb}^{\rm fit}}
\newcommand{\mtbw}{M_{tb}^{\rm w}}

\newcommand{\stp}{{\tilde{t}_1}}

\newcommand{\sbt}{{\tilde{b}_1}}

\newcommand{\glu}{{\tilde{g}}}
\newcommand{\mglu}{m_{\tilde{g}}}

\newcommand{\nedge}{N_{\rm edge}}
\newcommand{\Bredge}{{\rm Br(edge)}}
\newcommand{\nall}{N_{\rm all}}
\newcommand{\nfit}{N_{\rm fit}}

\begin{document}
\mark{{Study of stop }{Mihoko M. Nojiri}}
\title{
Study of $\tilde{t}$ and $\tilde{b}$ at LHC
}

\author{Mihoko M. Nojiri}
\address{YITP, Kyoto University, Kyoto 606-8502, Japan}
\abstract{
In supersymmetric models a gluino can decay into
$tb\tilde{\chi}^{\pm}_1$ through a stop or a sbottom.  The decay chain
produces an edge structure in the $m_{tb}$ distribution.  Monte Carlo
simulation studies show that the end point and the edge height would
be measured at the CERN LHC by using a sideband subtraction technique. 
The stop and sbottom masses as well as their decay branching ratios
are constrained by the measurement. We study interpretations of the
measurement.}

\maketitle

The minimal supersymmetric standard model (MSSM) is one of the
promising extensions of the standard model. The model requires
superpartners of the standard model particles (sparticles), and the
large hadron collider (LHC) at CERN might confirm the existence of the
new particles \cite{TDR}.
Among the sparticles the third generation squarks, stops
($\tilde{t}_i$) and sbottoms ($\tilde{b}_i$) ($i=1,2$), get special
imprints from physics at the very high energy scale.  Even is the
scalar masses are universal at a high energy scale, the third
generation squarks are much lighter than the first and second
generation squarks due to the Yukawa running effect. On the other
hand, some SUSY breaking models have non-universal boundary conditions
for the third generation mass parameters at the GUT scale.
Measurement of the sparticle masses provides
a way to probe the origin of the SUSY breaking in nature. 
%The
%sparticle mass measurement at the LHC has been therefore extensively
%studied. 

At the LHC, we may be able to access the nature of the stop and sbottom
provided that they are lighter than the gluino ($\tilde{g}$). In that
case, they copiously arise from the gluino decay. The relevant decay
modes for $\tilde{b}_i$ ($i=1,2$), $\tilde{t}_1$, to charginos
$\tilde{\chi}_j^{\pm}$ ($j=1,2$) or neutralinos $\tilde{\chi}^0_j$
($j=1,2,3,4$) are (indices to distinguish a particle and
its anti-particle is suppressed unless otherwise stated), ${\rm
(I)}_j~\glu\rightarrow b\sbt
\rightarrow bb\tilde{\chi}^0_j
\ (\rightarrow bbl^+l^-\tilde{\chi}^0_1)$,
${\rm (II)}_j~\glu\rightarrow t\stp \rightarrow tt\tilde{\chi}^0_j$,
${\rm (III)}_j~\glu\rightarrow t\stp \rightarrow 
tb\tilde{\chi}^{\pm}_j$,
${\rm (III)}_{ij}~\glu\rightarrow
b \tilde{b}_i \rightarrow  b W\stp \rightarrow 
bbW\tilde{\chi}^{\pm}_j$,
${\rm (IV)}_{ij}~\glu\rightarrow b \tilde{b}_i\rightarrow tb 
\tilde{\chi}^{\pm}_j$.
In previous literatures~\cite{TDR,Hinchliffe:1996iu}, the lighter 
sbottom $\tilde{b}_1$
is often studied through the mode (I)$_2$, namely the
$bb\tilde{\chi}^0_2\rightarrow bbl^+l^-\tilde{\chi}^0_1$
channel. This mode is important when the second lightest neutralino 
$\tilde{\chi}^0_2$ has
substantial branching ratios into leptons. 

Recently we proposed to measure the
edge position of the $m_{tb}$ distribution for the modes (III)$_{1}$
and (IV)$_{11}$, where $m_{tb}$ is the invariant mass of a top-bottom
($tb$) system\cite{Hisano:2002xq,Hisano:2003qu}.  
We focused on the reconstruction of hadronic decays of
the top quark, because the $m_{tb}$ distribution of the decay makes a
clear ``edge'' in this case.  The parton level $m_{tb}$ distributions
for the modes (III)$_j$ and (IV)$_{ij}$ are expressed as functions of
$\mglu$, $m_{\tilde{t}_1}$, $m_{\tilde{b}_i}$, and the chargino mass
$m_{\tilde{\chi}_j^{\pm}}$: $d\Gamma/dm_{tb}\propto m_{tb}$. 

The events containing $tb$ are selected by requiring the following
condition in addition to the standard SUSY cuts: 1) Two and only two $b$
jets. 2) Jet pairs consistent with a hadronic $W$ boson decay, $\vert
m_{jj}-m_W\vert < 15$~GeV.  3) The invariant mass of the jet pair and
one of the $b$-jets, $m_{bjj}$, satisfies $\vert m_{bjj}-m_t
\vert<30$~GeV. The events after the selection contain misreconstructed
events.  We use a $W$ sideband method to estimate the background
distribution due to misreconstructed events. Monte Carlo simulations
show that the distribution of the signal modes (III) and (IV) after
subtracting the background is very close to the parton level
distribution.  The distribution is then fitted by a simple fitting
function described with the end point $\mtbfit$, the edge height $h$
per $\Delta m$ bin, and a smearing parameter.

The edge position (end point) of the $m_{tb}$ distribution $M_{tb}$ for
the modes (III)$_1$ and (IV)$_{11}$ are  sometimes very close.
When they are experimentally indistinguishable, it is convenient to define
a weighted mean of the end points;
\begin{eqnarray}
\mtbw=\frac{\Br({\rm III}) M_{tb}({\rm III})_1+
\Br({\rm IV})_{11}M_{tb}({\rm IV})_{11}}
{\Br({\rm III})+\Br({\rm IV})_{11}},
\label{mtbw}
\end{eqnarray}
where $\Br({\rm III})  \equiv  
\Br({\rm III})_1 +\Br({\rm III})_{11}+ \Br({\rm III})_{21}$.

\begin{figure}[ht]
\epsfxsize=5cm
\centerline{\epsfbox{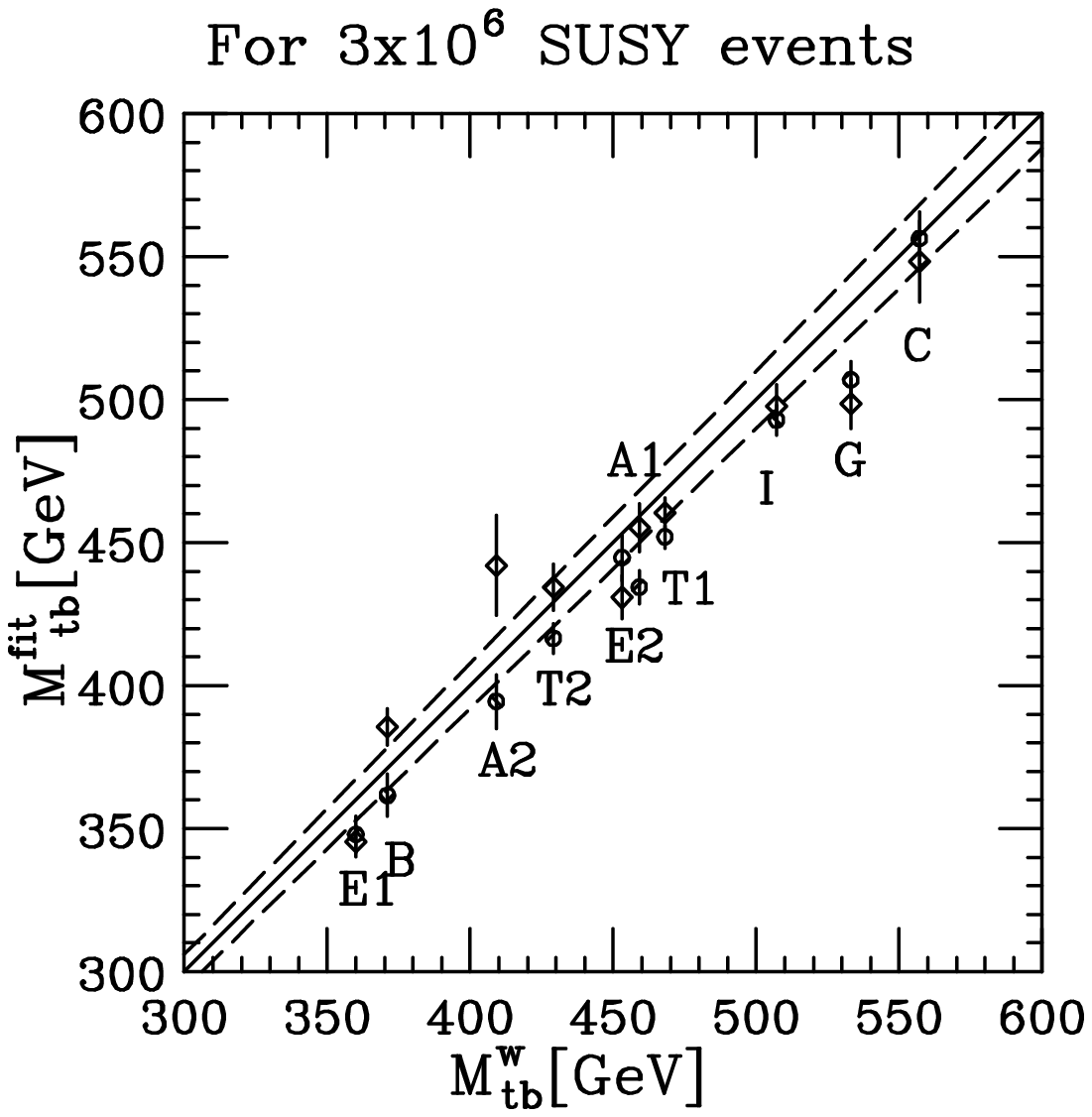} \hskip 1cm 
\epsfxsize=5cm  \epsfbox{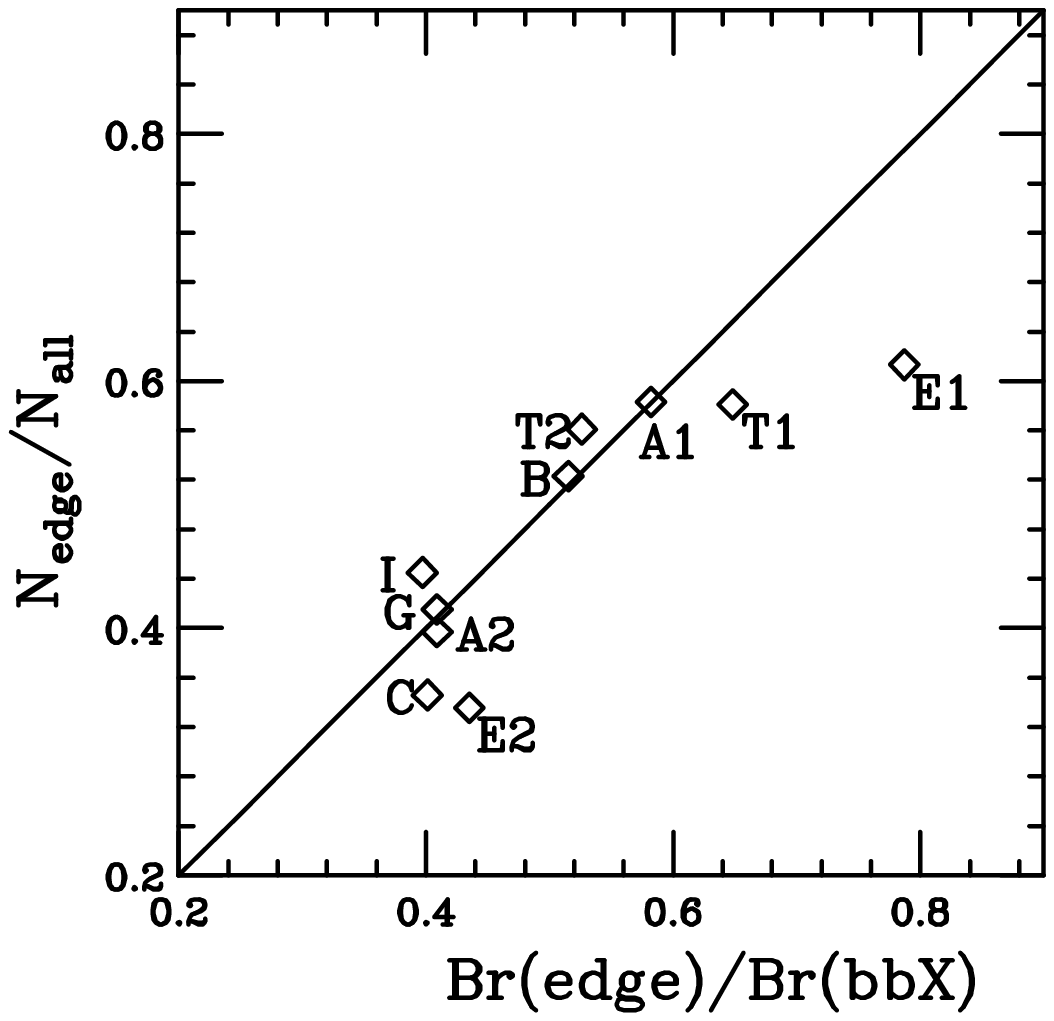}}
\caption{ 
Left:Relation between 
$M^{\rm w}_{tb}$ and 
$M^{\rm fit}_{tb}$ for the sample points. The solid line corresponds to  
$M^{\rm w}_{tb}= M^{\rm fit}_{tb}$. Bars with a diamond
and a circle correspond to  PYTHIA  and HERWIG samples, respectively. 
Right: Relation between
$\nedge/\nall$ and $\Bredge/\Br(\tilde{g}\rightarrow bbX)$. }
\label{fig:ratio}
\end{figure}

The relation between $\mtbw$ and $\mtbfit$ for several 
model points are shown in Fig. \ref{fig:ratio} (left). 
Here we generated $3\times
10^6$ SUSY events for each model point. The fitted value
$\mtbfit$ increases linearly with the weighted  end point $\mtbw$. 
The fitted value $\mtbfit$ tends to be lower 
than $\mtbw$, which is the effect of
particle missed outside the jet cones.  

We now discuss the relation between the edge height $h$ and the number
of reconstructed $tb$ events.  The number of the reconstructed
``edge'' events $\nedge$ arising from the decay chains (III) and (IV)
may be estimated from $M_{tb}^{\rm fit}$, $h$ per the bin size $\Delta
m$ as follows,
\begin{equation}
\nedge   \sim \nfit = 
\frac{h}{2}\left(\frac{m_t}{M^{\rm fit}_{tb}}  + 1\right)\times 
\frac{M^{\rm fit}_{tb}-m_{t}}{\Delta m }.
\label{nsig}\end{equation}
This formula is obtained by assuming the parton level distribution.
The consistency between $\nedge\sim\nfit$ is checked by using 
the generator information in Ref.~\cite{Hisano:2003qu}.

In the MSUGRA model, the decay modes which involve $W$ bosons (modes
(II), (III) and (IV)) often dominate the gluino decays to
$bbX$. In that case the
numbers of events  are given approximately as
\begin{eqnarray}
\nfit &\sim&  {\epsilon_{tb}}\Bredge 
 \left[2 N(\tilde{g}\tilde{g}) 
\left(1-\Br(\tilde{g}\rightarrow bbX)\right)+ N(\tilde{g}\tilde{q})
+ N(\tilde{g}\tilde{q}^*)\right],
\cr
\nall &\sim& {\epsilon_{tb}}\Br(\tilde{g}\rightarrow bbX)
\left[
2N(\tilde{g}\tilde{g}) 
\left(1-\Br(\tilde{g}\rightarrow bbX)\right)
+ N(\tilde{g}\tilde{q})+ N(\tilde{g}\tilde{q}^*)
\right],\cr&&
\end{eqnarray}
where
$\Bredge\equiv  \rm Br(III)_1 + Br(III)_{11}+ Br(III)_{21}
+ Br(IV)_{11}$
and  $\Br(\tilde{g}\to bbX)$ is the branching ratio of the gluino
decaying into stop or sbottom, thus having two bottom quarks in the
final state.  
Therefore  $\Bredge$ $/\Br(\tilde{g}\rightarrow bbX)\sim \nfit/\nall$
is expected.

In Fig.~\ref{fig:ratio}(right), we plot the ratio
$\nedge/\nall$ as a function of $\Bredge$ $/\Br(\tilde{g}\rightarrow
bbX)$. 
The points  tend to be on the expected line 
$\nedge/\nall= \Bredge /\Br(\tilde{g}\rightarrow bbX)$. 
Some points in the plots are away from the line: The point
``C'' is off because the chargino has large branching ratios into
leptons. 
At the point ``T1'', the stop mass is
significantly light and 
$\tilde{t}_1\tilde{t}^{*}_1$ productions contributes to $\nall$. The
points ``E1'' and ``E2'' are significantly off because the first and the
second generation squarks dominantly decay into the gluino, and the 
events containing two bottom quarks are not dominant.
\footnote{The description of the points are found in \cite{Hisano:2003qu}}.
These exceptional cases will be easily distinguished by looking 
into the data from LHC or a proposed O(1)TeV linear collider.

The $\mtbfit$ contains information on 
$\tilde{t}_1$ as can be seen in Eq.(\ref{mtbw}). 
However, the definition also depends on the electroweak SUSY parameters 
and the SUSY parameters in the sbottom sector. Some model assumptions 
or  inputs 
from the other measurements are necessary. In MSUGRA, the $M_{tb}$
and $h$
measurement constrain the trilinear coupling at GUT scale $A_0$. 
$\Delta A_0\sim 50$ GeV for $m_{\tilde{g}}\sim m_{\tilde{q}}\sim
700$~GeV are found.  On the other hand, if a $e^+e^-$ collider 
at O(1TeV) is built, the measurements 
of the sparticle productions at the LC would constrain 
the electroweak SUSY parameters precisely. 
In that case  the sbottom and stop studies 
at LHC would be significantly improved so that 
stop and sbottom masses and their mixing angle are measured
\cite{LHCLC}.

\noindent
{\bf Acknowledgment:}
I thank  J.~Hisano and K.~Kawawagoe for collaboration, and
the ATLAS collaboration members for useful discussion. We
have made use of the physics analysis framework and tools which
are the result of collaboration-wide efforts.
This work is supported in part by the Grant-in-Aid for 
Science Research, Ministry
of Education, Science and Culture, Japan 
(No.13135297 and  No.14046225 for JH,
No.11207101 for KK, and 
No.14540260 and 14046210 for MMN).

\end{document}